# A Survey of Mobile WiMAX IEEE 802.16m Standard.

Mr. Jha Rakesh
Deptt. Of E & T.C.
SVNIT
Surat, India
Jharakesh.45@gmail.com

Mr. Wankhede Vishal A.
Deptt. Of E & T.C.
SVNIT
Surat, India
wankhedeva@gmail.com

Prof. Dr. Upena Dalal
Deptt. Of E & T.C.
SVNIT
Surat, India
upena_dalal@yahoo.com

*Abstract*— **IEEE 802.16m amends the IEEE 802.16 Wireless MAN-OFDMA specification to provide an advanced air interface for operation in licenced bands. It will meet the cellular layer requirements of IMT-Advanced next generation mobile networks. It will be designed to provide significantly improved performance compared to other high rate broadband cellular network systems. For the next generation mobile networks, it is important to consider increasing peak, sustained data reates, corresponding spectral efficiencies, system capacity and cell coverage as well as decreasing latency and providing QoS while carefully considering overall system complexity. In this paper we provide an overview of the state-of-the-art mobile WiMAX technology and its development. We focus our discussion on Physical Layer, MAC Layer, Schedular,QoS provisioning and mobile WiMAX specification.**

*Keywords-Mobile WiMAX; Physical Layer; MAC Layer; Schedular; Scalable OFDM.*

I. INTRODUCTION

IEEE 802.16, a solution to broadband wireless access (BWA) commonly known as Worldwide Interoperability for Microwave Access (WiMAX), is a recent wireless broadband standard that has promised high bandwidth over long-range transmission. The standard specifies the air interface, including the medium access control (MAC) and physical (PHY) layers, of BWA. The key development in the PHY layer includes orthogonal frequency-division multiplexing (OFDM), in which multiple access is achieved by assigning a subset of subcarriers to each individual user [1]. This resembles code-division multiple access (CDMA) spread spectrum in that it can provide different quality of service (QoS) for each user; users achieve different data rates by assigning different code spreading factors or different numbers of spreading codes. In an OFDM system, the data is divided into multiple parallel substreams at a reduced data rate, and each is modulated and transmitted on a separate orthogonal subcarrier. This increases symbol duration and improves system robustness. OFDM is achieved by providing multiplexing on user's data streams on both uplink and downlink transmissions.

Lack of mobility support seems to be one of the major hindrances to its deployment compared to other standards such as IEEE 802.11 WLAN, since mobility support is widely considered as one of the key features in wireless networks. It is natural that the new IEEE 802.16e released earlier this year has added mobility support. This is generally referred to as mobile WiMAX [1].

Mobile WiMAX adds significant enhancements:

• It improves NLOS coverage by utilizing advanced antenna diversity schemes and hybrid automatic repeat request (HARQ).

• It adopts dense subchannelization, thus increasing system gain and improving indoor penetration.

• It uses adaptive antenna system (AAS) and multiple input multiple output (MIMO) technologies to improve coverage [2].

• It introduces a downlink subchannelization scheme, enabling better coverage and capacity trade-off [3-4].

This paper provides an overview of Mobile WiMAX standards and highlights potential problems arising from applications. Our main focuses are on the PHY layer, MAC layer specifications of mobile WiMAX. We give an overview of the MAC specification in the IEEE 802.16j and IEEE802.16m standards, specifically focusing the discussion on scheduling mechanisms and QoS provisioning. We review the new features in mobile WiMAX, including mobility support, handoff, and multicast services. We discuss technical challenges in mobile WiMAX deployment. We then conclude the paper.

II. PHYSICAL LAYER OF IEEE 802.16M.

This section contains an overview of some Physical Layer enhancements that are currently being considered for inclusion in future systems. Because the development of the 802.16m standard is still in a relatively early stage, the focus is on presenting the concepts and the principles on which the proposed enhancements will be based, rather than on providing specific implementation details. Although the exact degree of sophistication of the new additions to the standard cannot be safely predicted, it is expected that the additions will make some use of the concepts described below.

*A. Flexibility enhancements to support heterogeneous users in IEEE 802.16m:*

Because the goal of future wireless systems is to cater to needs of different users, efficient and flexible designs are needed. For some users (such as streaming low-rate applications) link reliability may be more important than high data rates, whereas others may be interested in achieving the maximum data rate even if a retransmission, and, therefore, additional delay may be required [4-6]. Moreover, the co-existence of different users should be achieved with relatively low control overhead. For these reasons, the frame format, the subcarrier mapping schemes and the pilot structure are being modified for 802.16m with respect to 802.16e. Each 802.16e frame consists of a downlink (DL) and an uplink (UL) part separated in time by an OFDMA symbol and is of variable size [3,7]. The (downlink or uplink) frame begins by control information that all users employ to synchronize and to determine if and when they should receive or transmit in the given frame. Control information is followed by data transmission by the base station (in the downlink subframe) or the mobile stations (in the uplink subframe). For each mobile station, transmission or reception happens in blocks that are constructed from basic units called slots. Each slot can be thought of as a two-dimensional block, one dimension being the time, the other dimension being the frequency. In general, a slot extends over one subchannel in the frequency direction and over 1 to 3 OFDMA symbols in the time direction, depending on the permutation scheme. The subchannels are groups of OFDMA subcarriers. The number of subcarriers per subchannel and the distribution of the subcarriers that make up a subchannel in the OFDMA symbol are determined based on the permutation scheme. As explained in more detail below, the subcarriers of a given subchannel are not always consecutive in frequency. Downlink and uplink subframes can be divided into different zones where different permutation schemes are used [9-10].

In the Partial Usage of Subchannels (PUSC) zone that is mandatory, the priority is to improve diversity and to spread out the effect of inter-cell interference. Each slot extends over 2 OFDMA symbols, and a subchannel consists of 24 data subcarriers that are distributed over the entire signal bandwidth (OFDMA symbol). Thus, each subchannel has approximately the same channel quality in terms of the channel gain and the inter-cell interference. To reduce the effect of the inter-cell interference, when PUSC is used, the available subchannels are distributed among base stations so that adjacent base stations not use the same subchannels. When the inter-cell interference is not significant, as in the case of mobile stations located closely to a base station, it may be advantageous to employ Full Usage of Subchannels (FUSC). The goal of the FUSC permutation scheme is similar to PUSC, i.e, to improve diversity and to spread out the effect of inter-cell interference. However, as the name suggests, in the FUSC zone all subchannels are used by a base station. For this reason, the design of the pilot pattern for the FUSC zone is slightly more efficient compared to PUSC. A subchannel in the FUSC permutation zone consists of 48 data subcarriers and the slot only comprises one OFDMA symbol.

### B. Extending use of MIMO transmission

Multiple-Input Multiple-Output (MIMO) communication is already a reality in wireless systems. It will be supported by the IEEE 802.11n amendment to the 802.11 WLAN standards that is expected to be ratified in the near future. Similarly, 802.16e includes support for MIMO downlink and uplink transmission. As MIMO technology matures and implementation issues are being resolved, it is expected that MIMO will be widely used for wireless communication. Current Mobile WiMAX profiles include support for up to 2 transmit antennas even though the IEEE 802.16e standard does not restrict the number of antennas, and allows up to 4 spatial streams. The current aim for Next Generation WiMAX systems is to support at least up to 8 transmit antennas at the base station, 4 streams and Space-Time Coding [2]. Moreover, although some other MIMO features of 802.16e, such as closed-loop MIMO, have not appeared in Mobile WiMAX profiles yet, it is expected that they will be included in new 802.16m-based systems. More specifically, it has been already decided to support closed-loop MIMO using Channel Quality Information, Precoding Matrix Index and rank feedback in future systems.

In 802.11 systems, as well as in the 802.16e standard, MIMO transmission is used to increase the data rate of the communication between a given transmitter-receiver pair and/or improve the reliability of the link. It is expected that 802.16m and future 3GPP systems will extend MIMO support to Multi-user (MU-) MIMO. More specifically, use of multiple antennas can improve the achievable rates of users in a network with given frequency resources. In information theoretic terms, the capacity region of the uplink and the downlink increases, in general, when MIMO transmission is employed [2]. In many cases, a large portion of this capacity increase can be achieved using relatively simple linear schemes (transmit beamforming at the downlink and linear equalizers at the uplink). Therefore, the achievable rates can be increased without the need for sophisticated channel coding. If larger complexity can be afforded, even higher gains can be attained using successive decoding at the uplink and Dirty Paper Coding schemes at the downlink. An overview of the projected MIMO architecture for the downlink of 802.16m systems is given in the System Description Document (SDD), and is repeated in Fig. 1 for convenience.

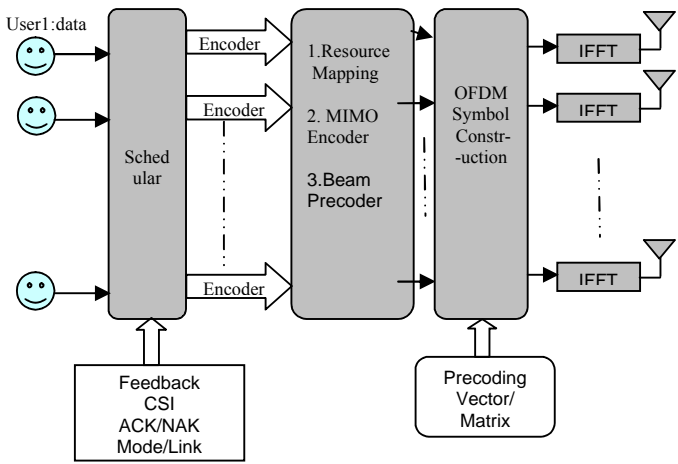

Figure 1. MIMO architecture for the downlink of 802.16m systems.

WiMAX and 3GPP networks employing MU-MIMO will need to calculate which users should transmit and receive during each frame, as well as the best achievable rate that corresponds to each user based on their QoS requirements, the number of users in each cell and their position. Although the information-theoretic capacity has been characterized, this is not an easy task, even for narrowband systems, and it is even more challenging when all subcarriers of the OFDMA system are considered. Therefore, efficient algorithms will be needed at the base station for user selection that will also determine the beamforming filters for the downlink, the receiver filters for the uplink and the required power allocation at the base station and each mobile station.

TABLE I. MOST IMPORTANT FEATURES AND SYSTEM REQUIREMENTS OF MOBILE WIMAX STANDARDS

| Requirement | IEEE 802.16e | IEEE802.16m |
|---|---|---|
| Aggregate Data Rate | 63 Mbps | 100 Mbps for mobile stations, 1 Gbps for fixed |
| Operating Radio Frequency | 2.3 GHz, 2.5-2.7 GHz, 3.5 GHz | < 6 GHz |
| Duplexing Schemes | TDD and FDD | TDD and FDD |
| MIMO support | up to 4 streams, no limit on antennas | 4 or 8 streams, no limit on antennas |
| Coverage | 10 km | 3 km, 5-30 km and 30-100 km |
| Handover Inter-frequency Interruption Time | 35-50 ms | depending on scenario |
| Handover Intra-frequency Interruption Time | Not Specified | 30 ms |
| Handover between 802.16 standards (for corresponding mobile station) | From 802.16e serving BS to 802.16e target BS | 100 ms |
| Handover with other technologies | Not Specified | From legacy serving BS to legacy target BS From 802.16m serving BS to legacy target BS From legacy serving BS to 802.16m target BS From 802.16m serving BS to 802.16m target BS |
| Mobility Speed | Vehicular: 120 km/h | IEEE 802.11, 3GPP2, GSM/EDGE, (E-)UTRA (LTE TDD) Using IEEE 802.21 Media Independent Handover (MIH) |
| Position accuracy | Not Specified | Indoor: 10 km/h Basic Coverage Urban: 120 km/h High Speed: 350 km/h Location Determination Latency: 30 s |

### C. Resource allocation and multi-cell MIMO

In cellular networks careful frequency planning is required in order to achieve communication with small outage probability and, at the same time, minimize interference among users of neighboring cells. Users near the cell edges are particularly vulnerable, because they receive signals of comparable strength from more than one base stations [2]. For this reason, different parts of the frequency spectrum are typically assigned to neighboring cells. The assignment in current systems is static and can only be changed by manual re-configuration of the system. Changes to the frequency allocation can only be performed periodically and careful cell planning is required in order not to affect other parts of the system. Frequencies are reused by cells that are sufficiently far away so that the interference caused by transmissions on the same frequencies is small enough to guarantee satisfactory Signal- to-Interference and Noise Ratios (SINRs). Although static frequency reuse schemes greatly simplify the design of cellular systems, they incur loss in efficiency because parts of the spectrum in some cells may remain unused while, at the same time, other cells may be restricting the rates of their mobile stations or even denying admission to new users. Moreover, the handover process is more complicated for mobile stations since communication in more than one frequencies is required.

### D. Interoperability and coexistence.

In order for the standard to be able to support either legacy base and mobile stations or other technologies (e.g. LTE), the concept of the time zone, an integer number (greater than 0) of consecutive subframes, is introduced.

Interoperability among IEEE 802.16 standards [11]: The 802.16m Network Reference Model permits interoperability of IEEE 802.16m Layer 1 and Layer 2 with legacy 802.16 standards. The motivation for ensuring interoperability comes from the fact that WiMAX networks have already been deployed, and it is more realistic to require interoperability instead of an update of the entire network. Another advantage is that each 802.16 standard provides specific functionalities in a WiMAX network. The goal in 802.16m is to enable coexistence of all these functionalities in a network without the need to create a new standard that contains all of them. The supported connections and frame structure are summarized in Fig. 2 and Fig. 3. The legacy standard can transmit during the legacy zones (also called LZones), whereas 802.16m-capable stations can transmit during the new zones. The Uplink (UL) portion shall start with the legacy UL zone, because legacy base stations, mobile stations or relays expect IEEE 802.16e UL control information to be sent in this region. When no stations using a legacy 802.16 standard are present, the corresponding zone is removed. The zones are multiplexed using TDM in the downlink, whereas both TDM and FDM can used in the uplink. In each connection, the standard that is in charge is showcased. The Access Service Network can be connected with other network infrastructures (e.g. 802.11, 3GPP etc.) or to the Connectivity Service Network in order to provide Internet to the clients.

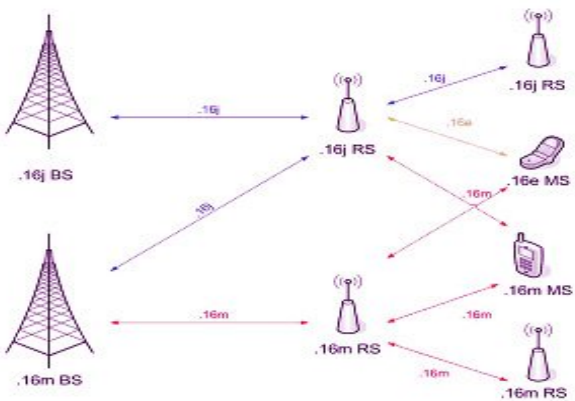

Figure 2. Supported 802.16 connections

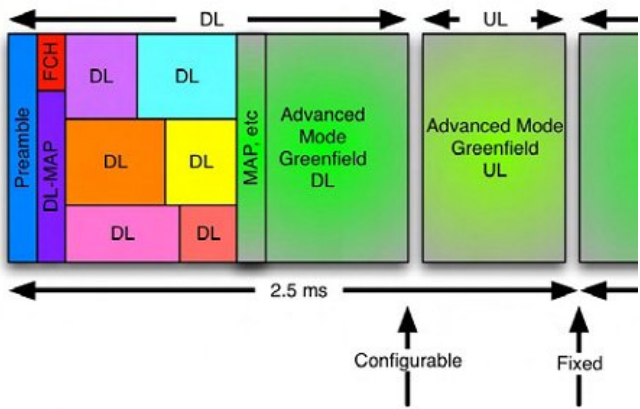

Figure 3. IEEE 802.16m frame structure with TDM Downlink and FDM Uplink

### III. BASIC FUNCTIONALITY OF MAC LAYER IN WIMAX

Figure 4 presents the reference model in IEEE 802.16. The MAC layer consists of three sublayers: the service-specific convergence sublayer (CS), MAC common part sublayer (MAC CPS), and security sublayer. The main functionality of the CS is to transform or map external data from the upper layers into appropriate MAC service data units (SDUs) for the MAC CPS. This includes classification of external data with the proper MAC service flow identifier (SFID) and connection identifier (CID). An SDU is the basic data unit exchanged between two adjacent protocol layers. [11,14] The MAC CPS provides the core functionality for system access, allocation of bandwidth, and connection establishment and maintenance. This sublayer also handles the QoS aspect of data transmission. The security sublayer provides functionalities such as authentication, secure key exchange, and encryption. For the PHY layer, the standard supports multiple PHY specifications, each handling a particular frequency range. The MAC CPS contains the essential functionalities for scheduling and QoS provisioning in the system.

IEEE 802.16d MAC provides two modes of operation: point-to-multipoint (PMP) and multipoint-to-multipoint (mesh) [13]. The functionalities of the MAC sublayer are related to PHY control (cross-layer functionalities, such as HARQ ACK/NACK etc). The Control Signaling block is responsible for allocating resources by exchanging messages such as DL-MAP and UL-MAP. The QoS block allocates the input traffic to different traffic classes based on the scheduling and resource block, according to the SLA guarantees. The name of other blocks, such as fragmentation/packing, multi-radio coexistence and MAC PDU formation, clearly describes their function. The MAC sublayer also deploys state-of-the-art power saving and handover mechanisms in order to enable mobility and make connections available to speeds up to 350 km/h. Since newer mobile devices tend to incorporate an increasing number of functionalities, in WiMAX networks the power saving implementation incorporates service differentiation on power classes. A natural consequence of any sleeping mechanism is the increase of the delay. Thus, delay-prone and non delay-prone applications are allocated to different classes, such that the energy savings be optimized, while satisfying the appropriate QoS (e.g those that support web page downloading or emails). MAC addresses play the role of identification of individual stations. IEEE 802.16m introduces two different types of addresses in the MAC sublayer. 1) The IEEE 802 MAC address that has the generic 48-bit format and 2) two MAC logical addresses that are assigned to the mobile station by management messages from the base station. These addresses are used for resource allocation and management of the mobile station and are called "Station Identifiers" (assigned during network entry) and "Flow Identifiers" (assigned for QoS purposes).

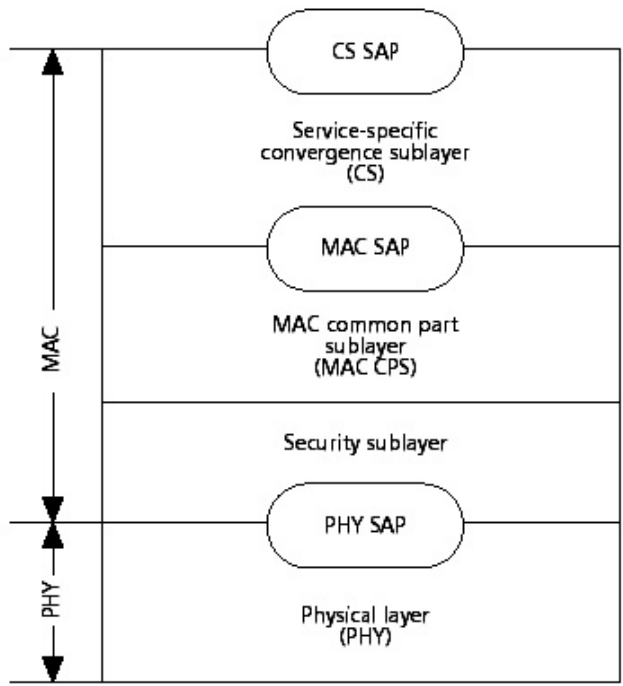

Figure 4. IEEE 802.16 reference model.

## IV. SCHEDULER

Scheduling is the main component of the MAC layer that helps assure QoS to various service classes [12,13,14,16]. The scheduler works as a distributor to allocate the resources among MSs. The allocated resource can be defined as the number of slots and then these sots are mapped into a number of subchannels (each subchannel is a group of multiple physical subcarriers) and time duration (OFDM symbols). In OFDMA, the smallest logical unit for bandwidth allocation is a slot. The definition of slot depends upon the direction of traffic (downlink/uplink) and subchannelization modes. For example, in PUSC mode in downlink, one slot is equal to twenty four subcarriers (one subchannel) for three OFDM symbols duration. In the same mode for uplink, one slot is fourteen subcarriers (one uplink subchannel) for two OFDM symbols duration. The mapping process from logical subchannel to multiple physical subcarriers is called a permutation. PUSC, discussed above is one of the permutation modes. Others include Fully Used Subchannelization (FUSC) and Adaptive Modulation and Coding (band-AMC). The term band-AMC distinguishes the permutation from adaptive modulation and coding (AMC) MCS selection procedure. Basically there are two types of permutations: distributed and adjacent. The distributed subcarrier permutation is suitable for mobile users while adjacent permutation is for fixed (stationary) users. After the scheduler logically assigns the resource in terms of number of slots, it may also have to consider the physical allocation, e.g., the subcarrier allocation. In systems with Single Carrier PHY, the scheduler assigns the entire frequency channel to a MS. Therefore, the main task is to decide how to allocate the number of slots in a frame for each user. In systems with OFDM PHY, the scheduler considers the modulation schemes for various subcarriers and decides the number of slots allocated. In systems with OFDMA PHY, the scheduler needs to take into consideration the fact that a subset of subcarriers is assigned to each user. Scheduler designers need to consider the allocations logically and physically. Logically, the scheduler should calculate the number of slots based on QoS service classes. Physically, the scheduler needs to select which subchannels and time intervals are suitable for each user. The goal is to minimize power consumption, to minimize bit error rate and to maximize the total throughput. There are three distinct scheduling processes: two at the BS - one for downlink and the other for uplink and one at the MS for uplink as shown in Fig. 5. At the BS, packets from the upper layer are put into different queues, which ideally is per-CID queue in order to prevent head of line (HOL) blocking. However, the optimization of queue can be done and the number of required queues can be reduced. Then, based on the QoS parameters and some extra information such as the channel state condition, the DL-BS scheduler decides which queue to service and how many service data units (SDUs) should be transmitted to the MSs. Since the BS controls the access to the medium, the second scheduler - the UL-BS scheduler - makes the allocation decision based on the bandwidth requests from the MSs and the associated QoS parameters. Several ways to send bandwidth requests were described earlier in Section I.F. Finally, the third scheduler is at the MS. Once the UL-BS grants the bandwidth for the MS, the MS scheduler decides which queues should use that allocation. Recall that while the requests are per connections, the grants are per subscriber and the subscriber is free to choose the appropriate queue to service. The MS scheduler needs a mechanism to allocate the bandwidth in an efficient way. Fig. 6 classification of scheduler is given.

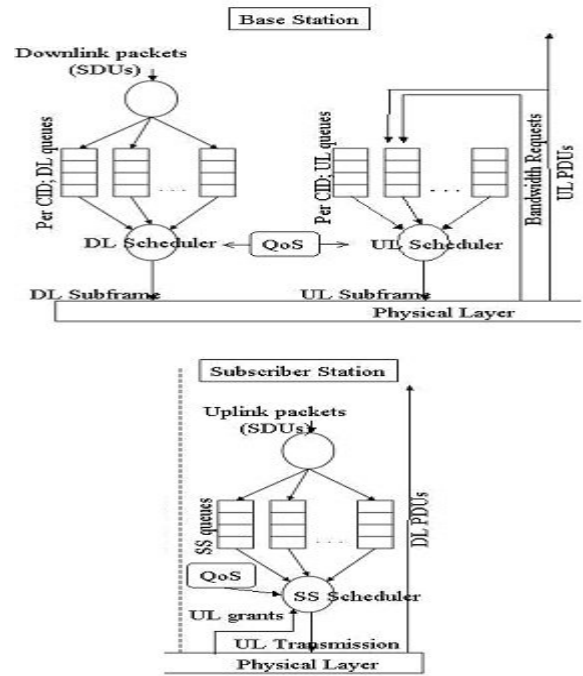

Figure 5. Component Schedulers at BS and MSs

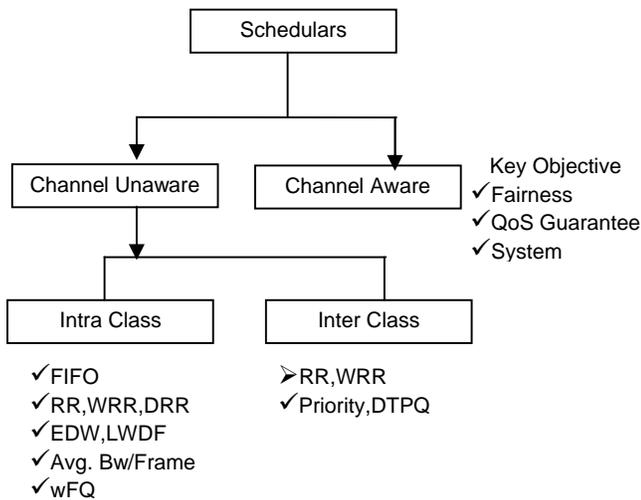

Figure. 6. Classification of WiMAX schedulers

## V. WiMAX QoS Service Classes

IEEE 802.16 defines five QoS service classes: Unsolicited Grant Scheme (UGS), Extended Real Time Polling Service (ertPS), Real Time Polling Service (rtPS), Non Real Time Polling Service (nrtPS) and Best Effort Service (BE). Each of these has its own QoS parameters such as minimum throughput requirement and delay/jitter constraints. Table II presents a comparison of these classes [15-16].

UGS: This service class provides a fixed periodic bandwidth allocation. Once the connection is setup, there is no need to send any other requests. This service is designed for constant bit rate (CBR) real-time traffic such as E1/T1 circuit emulation. The main QoS parameters are maximum sustained rate (MST), maximum latency and tolerated jitter (the maximum delay variation).

ertPS: This service is designed to support VoIP with silence suppression. No traffic is sent during silent periods. ertPS service is similar to UGS in that the BS allocates the maximum sustained rate in active mode, but no bandwidth is allocated during the silent period. There is a need to have the BS poll the MS during the silent period to determine if the silent period has ended. The QoS parameters are the same as those in UGS.

rtPS: This service class is for variable bit rate (VBR) realtime traffic such as MPEG compressed video. Unlike UGS, rtPS bandwidth requirements vary and so the BS needs to regularly poll each MS to determine what allocations need to be made. The QoS parameters are similar to the UGS but minimum reserved traffic rate and maximum sustained traffic rate need to be specified separately. For UGS and ertPS services, these two parameters are the same, if present.

nrtPS: This service class is for non-real-time VBR traffic with no delay guarantee. Only minimum rate is guaranteed. File Transfer Protocol (FTP) traffic is an example of applications using this service class.

Most of data traffic falls into this category. This service class guarantees neither delay nor throughput. The bandwidth will be granted to the MS if and only if there is a left-over bandwidth from other classes. In practice most implementations allow specifying minimum reserved traffic rate and maximum sustained traffic rate even for this class.

Note that for non-real-time traffic, traffic priority is also one the QoS parameters that can differentiate among different connections or subscribers within the same service class. Consider bandwidth request mechanisms for uplink. UGS, ertPS and rtPS are real-time traffic. UGS has a static allocation. ertPS is a combination of UGS and rtPS. Both UGS and ertPS can reserve the bandwidth during setup. Unlike UGS, ertPS allows all kinds of bandwidth request including contention resolution. rtPS can not participate in contention resolution. For other traffic classes (non real-time traffic), nrtPS and BE, several types of bandwidth requests are allowed such as piggybacking, bandwidth stealing, unicast polling and contention resolution. These are further discussed in Section I.F. Thus mobile WiMAX brings potential benefits in terms of coverage, power consumption, self-installation, frequency reuse, and bandwidth efficiency. One of the key complications is that the incompatibility in the newly introduced scalable OFDM (SOFDM) in IEEE 802.11e with the original OFDM scheme forces equipment manufacturers to come up with mechanisms to ease the transition

TABLE II. COMPARISON OF WIMAX QOSSERVICE CLASSES

| QoS | Pros | Cons |
|---|---|---|
| UGS | No overhead. Meet guaranteed latency for real- time service | Bandwidth may not be utilized fully since allocations are granted regardless of current need |
| ertPS | Optimal latency and data overhead efficiency | Need to use the polling mechanism(to meet the delay guarantee) and a mechanism to let the BS know when the traffic starts during silent perios |
| rtPS | Optimal data transport efficiency | Require the overhead of bandwidth request and the polling latency(to meet the delay guarantee) |
| nrtPS | Provide efficient service for non-real-time traffic with minimum reserved rate | N/A |
| BE | Provide efficient service for BE traffic | No service guarantee, some connections may starve for long period of time |

## VI. CONCLUSION

This paper presents an overview of the IEEE 802.16m PHY layer issues ,MAC protocol, specifically issues associated with scheduling and QoS provisioning. It also discusses the main features of the newly standardized mobile WiMAX, IEEE 802.16e to IEEE 802.16m. With the introduction of mobile WiMAX technology, it can be expected that future work will focus on the mobility aspect and interoperability of mobile WiMAX with other wireless technologies. For high quality voice and video, Internet and mobility, demand for bandwidth is more. To address these needs IEEE 802.16m appears as a strong candidate for providing aggregate rates to high-speed mobile users at the range of Gbps.